\documentclass[twocolumns]{nature}
\usepackage{color} 
\usepackage{graphicx} 
\usepackage[usenames,dvipsnames]{xcolor} 
\usepackage{amsmath,amssymb}
\usepackage{url}
\usepackage{pdfpages}

\title{Phenomenological theory of collective decision-making}

\author{Anna Zafeiris$^1$, Zsombor Koman$^2$, Enys Mones$^3$ \& Tam\'{a}s Vicsek$^{1,2}$}

\begin{document}

\maketitle

\begin{affiliations}
 \item Statistical and Biological Physics Research Group of HAS, P\'{a}zm\'{a}ny P\'{e}ter s\'{e}t\'{a}ny 1A, H-1117, Budapest, Hungary
 \item Department of Biological Physics, E\"{o}tv\"{o}s University, P\'{a}zm\'{a}ny P\'{e}ter s\'{e}t\'{a}ny 1A, H-1117, Budapest, Hungary
 \item Department of Applied Mathematics and Computer Science, Technical University of Denmark, 2800 Kgs. Lyngby, Denmark
\end{affiliations}

\begin{abstract}
An essential task of groups is to provide efficient solutions for the complex problems they face. Indeed, considerable efforts have been devoted to the question of collective decision-making related to problems involving a single dominant feature. Here we introduce a quantitative formalism for finding the optimal distribution of the group members' competences in the more typical case when the underlying problem is complex, i.e., multidimensional. Thus, we consider teams that are aiming at obtaining the best possible answer to a problem having a number of independent sub-problems. Our approach is based on a generic scheme for the process of evaluating the proposed solutions (i.e., negotiation). We demonstrate that the best performing groups have at least one specialist for each sub-problem -- but a far less intuitive result is that finding the optimal solution by the interacting group members requires that the specialists also have some insight into the sub-problems beyond their unique field(s). We present empirical results obtained by using a large-scale database of citations being in good agreement with the above theory. The framework we have developed can easily be adapted to a variety of realistic situations since taking into account the weights of the sub-problems, the opinions or the relations of the group is straightforward. Consequently, our method can be used in several contexts, especially when the optimal composition of a group of decision-makers is designed.
\end{abstract}

\section*{Introduction}
Addressing the process of collective decision making has represented a great scientific challenge for a long time \cite{forsyth2006effective,surowiecki2005wisdom,clearwater1991cooperative,planas2015group}. It is a highly relevant aspect of the behavior of social groups, in particular, because as it has been argued, measured and shown analytically: the ``wisdom of crowds'' can go qualitatively beyond that of the individuals' \cite{surowiecki2005wisdom}. This statement also holds for animal assemblies \cite{conradt2009group,nagy2010hierarchical,couzin2011uninformed}. A rarely considered, but essential case is when the problem to be solved is complex, i.e., has many facets. Under such conditions the quality of the collective solution is highly influenced by the composition of the group. Obviously, if the members of the group are identical, the group's performance can hardly go beyond that of any of its member's. However, if the problem to be solved is complex -- i.e., has a number of different aspects or ``dimensions'' \cite{vicsek2002complexity} -- a group having members specialized in their respective kinds of sub-problems is expected to be much more efficient in providing a high quality answer, than a uniform one. The stress is on the independent nature of the sub-problems, making the problem high-dimensional. In a way our present work can be considered as a quantitative approach to the problem of division of labor \cite{smith1838inquiry,durkheim2014division} in the context of collective decision making (the task/labor is to bring about a decision; the division is made among the specialists of the sub-problems).

In spite of the above almost trivial observation regarding heterogeneous, diverse or  ``multidimensional'' groups, a quantitative demonstration of its validity needs a carefully constructed framework. Prior works involving quantitative analysis have almost exclusively focused on problems that could be regarded as ``one-dimensional'' \cite{surowiecki2005wisdom,page2010diversity,hong2001problem,zafeiris2013group} from our point of view which considers a problem having several dimensions (being multidimensional) if it can be broken down into sub-problems, each having its own characteristic feature independent of those of the others'. In the case of one-dimensional problems it has been demonstrated -- using approaches from theory (see, eg., the pioneering works \cite{page2010diversity,zafeiris2013group}) through genetic optimization \cite{zafeiris2013group} to agent based modeling/simulations \cite{guttal2010social} and observations \cite{hamilton2003team,ruderman1996selected} -- that diverse groups can outperform homogeneous ones. 
 
Intuition suggests that a group of specialists (one competent person for each sub-problem) should be optimal regarding the quality of the solution with the constraint of minimizing costs at the same time. Here we present a generic agent-based approach which -- due to its minimal assumptions -- quantitatively demonstrates that the breadth of knowledge of its members makes a group more efficient, i.e., being capable of using a smaller amount of resources to produce a more beneficial solution in a wide variety of potential applications. This is what corresponds to the ``synergy'' resulting in a better decision relative to the one following from a simple ``linear'' aggregation of the proposed solutions. And what we show in our work is how this synergy can emerge from a negotiation process. Naturally, negotiation is absent (generally) in animal societies. Specialization is the result of age or hormone level etc.

A paradigmatic example for our approach is that of a board of directors for a large company (however, there are many other possible examples ranging from a group of animals searching for resources up to a government or simply a team carrying out interdisciplinary research). In the case of a board of directors a potential candidate problem is that of finding the best possible placement and product for a new factory. Obviously, the various aspects of this problem are quite diverse, each of them requiring specific knowledge, i.e., the decision involves knowledge of the history of the given country, various features of the labor force (education, etc.), geographical and logistic conditions, potential market in the region, and so on. It is an important feature of the situation that the members of the group cannot get any information about the quality of their propositions from an ``outsider'' who could know the optimal solution \textit{ab ovo}.

\section*{Results}
\subsection{Modeling collective decision-making}
We have aimed at a model that is simple, but is still appropriate for projecting a wide class of realistic situations onto it. In our approach, the process of collective decision-making is divided into four basic stages (described in the third subsection).

\subsubsection*{i) Basic assumptions}
We consider groups of $N$ individuals solving a problem $P$ having $M$ sub-problems $P_j$ ($j=1,2,\mathrm{…},M$) such that for addressing each sub-problem a unique (specific) skill is needed. Thus, we are dealing with a set of $N \times M$ abilities or levels/degrees of skill, $A_{ij}$ ($i=1,2,\mathrm{…},N$), these values being proportional to the ability/competence (e.g., accuracy) of an individual $i$ to give the best answer for the $j$th sub-problem. We do not initially specify the $A_{ij}$ parameters: the method we apply will result in their optimal values.

\subsubsection*{ii) Defining optimal groups}
Without losing the generality of the above setting, we assume that $A_{ij}$-s take their values from the unit interval $[0, 1]$. The ability matrix is also related to the costs involved in finding a solution (since acquiring a high ability to successfully address a sub-problem involves costs, such as experience, learning, etc.). It is obvious that the cost of obtaining an ability $A$ is typically not a linear function of $A$, since achieving the capacity of perfect knowledge ($A=1$) is much more costly than achieving a partial knowledge (e.g., $A=0.5$). For the sake of simplicity we assume that the cost $C$ for obtaining ability $A$, is 
\begin{equation}
C = f(A)= \mathrm{const} \cdot A^x
\end{equation}
(where $1< x$, and $\mathrm{const}$ is a constant corresponding to the relative weights of the costs, when calculating the fitness of a group for given $A_{ij}$-s). We start with a random distribution of the $A_{ij}$ values and search for their optimal distribution (by letting them evolve). Here optimal distribution means one which provides the best possible solution for the smallest possible -- or for a given prefixed -- cost. 

\subsubsection*{iii) Opinions, information diffusion and decision-making}
During the ``information diffusion'' phase the members of the group interact through evaluating the solutions of the sub-problems ``proposed'' by the other members such that the quality of a given proposition (by member $i$ for the $j$th problem) is (in the simplest case) assumed to be 
\begin{equation}
Q_{ij}=A_{ij}.
\end{equation}

First, each member evaluates all the propositions. The evaluation made by member $i'$ regarding the quality $Q_{ij}$ is denoted by $E_{ij}^{i'}$ and it is proportional to both $Q_{ij}$ and $A_{i'j}$. The accuracy of such an evaluation is distorted by a stochastic factor representing that those members who have small abilities to evaluate a proposal tend to make mistakes in their appreciation with an amplitude involving randomness. These evaluations ($E_{ij}^{i'}$) represent the central ingredient of our approach.

These are next (in several rounds of an imaginary ``round table discussion'') modified by further interactions (communication/evaluation) with other group members, $i''$, chosen with a probability proportional to their abilities concerning problem $P_j$, $A_{i''j}$. The total number of additional evaluations in a given decision-making event is equal to $X\%$ of $N$. The quality of the solution for a given $P_j$ is obtained by accepting the proposal of member $i^*$ receiving the highest average evaluation 
\begin{equation}
E_j^\mathrm{max} = E_{i^*j},
\end{equation}
where
\begin{equation}
i^* = \mathrm{argmax}_i E_{ij} = \mathrm{argmax}_i \langle E_{ij}^{i^\prime}\rangle_{i'}.
\end{equation}
Here $\langle\cdot\rangle_{i'}$ denotes the average over $i'$.
By the other members concerning his/her proposition for the solution of problem $P_j$ i.e., 
\begin{equation}
Q_j^\mathrm{max} = Q_{i^*j} = A_{i^*j}.
\end{equation}

The quality $Q$ of the solution given for $P$ -- provided by the whole group -- is then obtained by aggregating the proposals having the highest evaluations for the $P_j$-s after the last round. A simplified flow diagram of how the model works is given in Fig.~\ref{fig:chart}. It is meant to be self-explanatory, more details are provided in the Appendix.

\begin{figure}
\centering
\includegraphics[width=.9\linewidth]{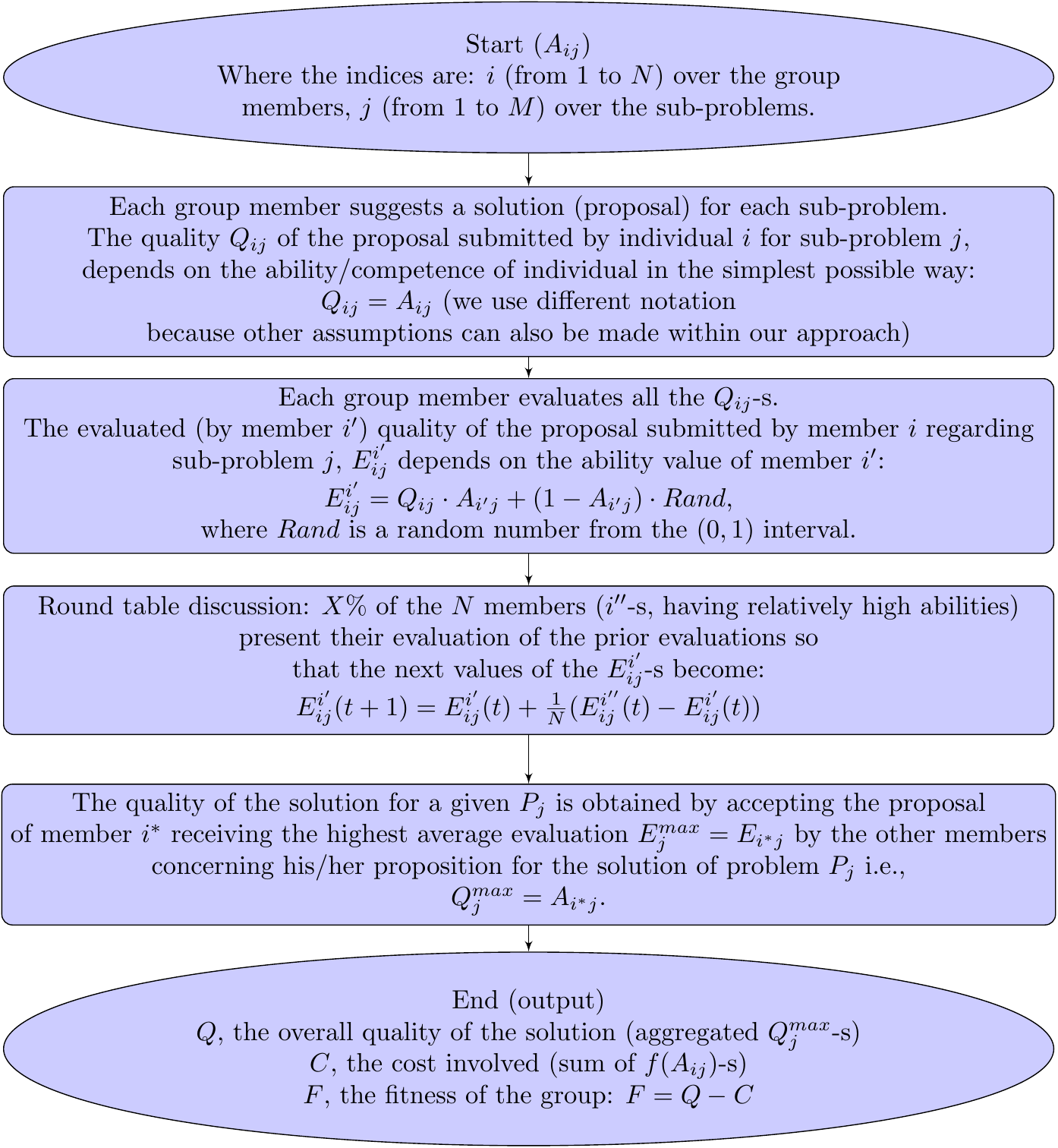}\\
\caption{Flow diagram representation of a single step (generation) during the genetic optimization. For notations see the text. This chart gives a detailed description of the process through which the corresponding fitness (efficiency) value $F$ is calculated based on the given ability matrix ($A_{ij}$). In the next step of the genetic algorithm these $F$-s are used as weights based on which the ``parents'' of the new generation are chosen (after the combination of two parents, random perturbations, ``mutations'' are applied before finalizing the groups of the ``young'' generation).}
\label{fig:chart}
\end{figure}

\section*{Results from simulations}
\begin{figure}
\centering
\includegraphics[width=.9\linewidth]{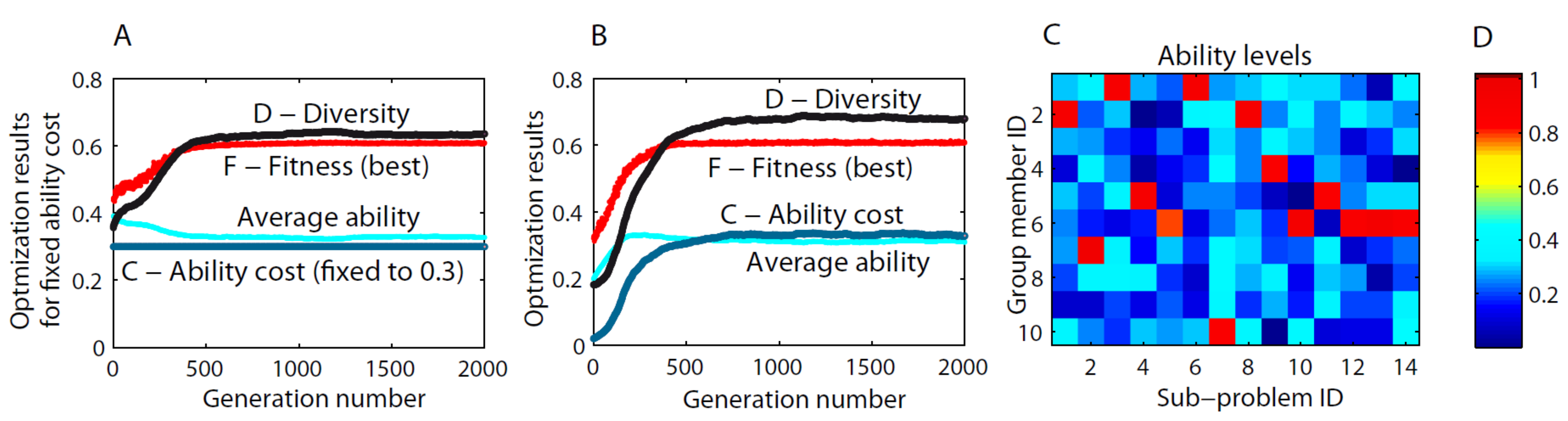}\\
\caption{Illustration of both the process (A,B) and the end result (C) of calculating the optimal distribution of abilities/competences, $A_{ij}^\mathrm{max}$, using a genetic optimization method. In (A) the generation number ($G$) dependence of the average fitness values ($F$) of the groups is plotted (red) for the fixed amount of cost, $C=0.3$ (dark blue). The averaging is made over a population size of 2000 groups. The corresponding diversity, $D$, is indicated by the black line. The groups had $N=10$ members and $M=14$ sub-problems had to be answered. In (B) we display the evolution of the relevant parameters when the optimization is done with non-fixed ability cost $C$. (C) Displays the optimal ability matrix visualized with colors -- the scale being indicated in (D). These results are for a generic case into which a few plausible assumptions are incorporated: the sub-problems have equal importance (weight) and $X=30\%$ of the members take role in the round-table discussion. The most relevant message of (C) is that there is one specialist for each sub-problem (not necessarily one person per sub-problem) and, perhaps rather intriguingly, the specialists are found to have a clearly non-negligible competence concerning several of the other sub-problems. If we add some cost for the case when a single person is a specialist of more than one sub-problem, the solution ceases to have multiple specialities per person.}
\label{fig:1}
\end{figure}

\paragraph*{General properties}
In Fig.~\ref{fig:1} we show results for $A_{ij}^\mathrm{max}$ using equation \ref{eq:fitn}, i.e., evaluating both the quality and the cost of the obtained $A_{ij}$ and considering the average of the entire population at the end of the evolutionary process. In most of the figures -- in addition to visualizing the values of $A_{ij}^\mathrm{max}$ -- we also plot how the fitness $F$, the quality of the solution $Q$ and the cost $C$ changes as a function of the generation number $G$ (as the population of groups evolves). In addition, we also display how the diversity $D$ of the abilities depends on $G$. In all cases we find that the optimal distribution of the abilities is highly diverse. In all plots we use $N=10$ and $M=14$ without loss of generality (the main features of the optimal ability distribution do not differ qualitatively for different $N$ and $M$ pairs).

Figure~\ref{fig:1} demonstrates some relevant features of both the process (the progress of the genetic algorithm) and the outcome of optimizing the ability distribution. Random initial conditions correspond to relatively low fitness and high costs. The efficiency/fitness of a group quickly increases at the first stage of the optimization. An important observation is that higher fitness is accompanied by larger diversity values ($D$). 

Our results come from simple and realistic assumptions regarding the ``negotiation/discussion'' process. Although the corresponding rules and calculations are not trivially transparent at all, nevertheless a relatively plausible interpretation for the main result can be provided. Perhaps the most essential step in our algorithm is the one when the group members, one after another, provide an evaluation of the proposals of the other members. If a member has zero ability to evaluate the proposal for a given sub-problem, then the contribution of this member to choosing the otherwise very good proposition becomes totally erratic (see Fig.~\ref{fig:chart}, defining equation of $E_{ij}^{i'}$). Conversely, even a relatively small ability to estimate the right value of a proposal results in a decreased level of randomness in the evaluation and, in this way, provides a more accurate estimated proposition quality. When the evaluations are aggregated to choose the best answer, the latter, more consistent contributions become to play an essential role.

\begin{figure}
\centering
\includegraphics[width=.99\linewidth]{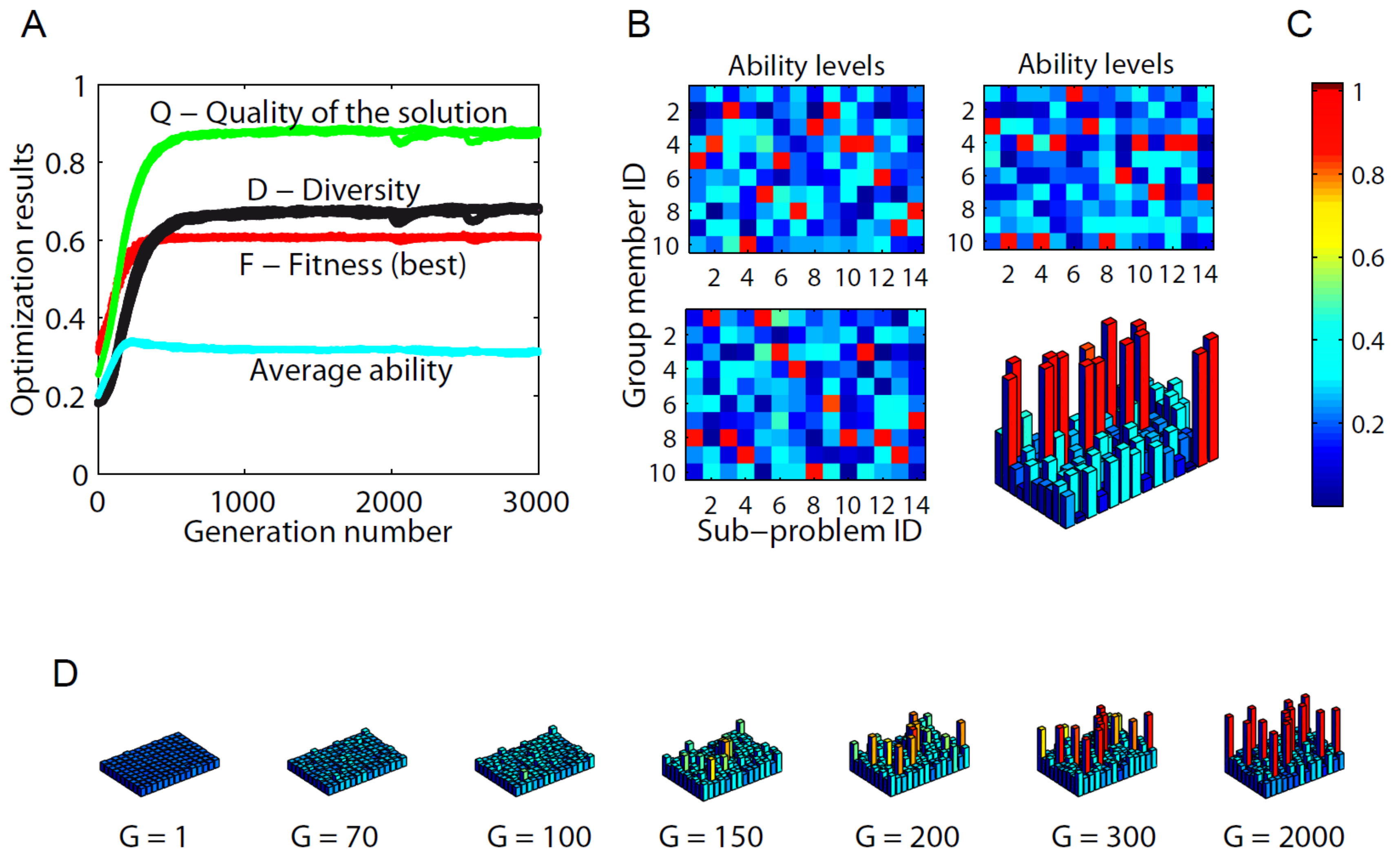}
\caption{Visualization of the optimization for four different (random) initial conditions and (stochastic) realizations. Although the individual ability distributions are different, they correspond to about the same level of optimality which can be seen from the four curves virtually overlapping in all cases. The wiggles around $G=2000$ and $G=2500$ are due to the momentarily increased level of perturbations or ``mutations'' within the genetic algorithm (in the spirit of simulated annealing, see Materials and methods). (D) displays the development of the ability matrix as the genetic algorithm progresses. Here and in one of the displays in (B) a combination of column heights and colors is used to visualize the values of the ability matrix.}
\label{fig:2}
\end{figure}

\paragraph*{Robustness}
Next we investigate the robustness of the new results stemming from our approach by testing the method on a few specific conditions. First, we start the optimization from different initial conditions and check whether the results are consistent with each other (have the same overall features). Figure~\ref{fig:2} shows two different directions of the comparison. In Fig.~\ref{fig:2}A we show how similarly the main quantities ($A$, $D$, $F$ and $Q$) evolve during four (stochastically) independent optimization processes starting from different random initial conditions and lead to rather different final configurations (presented in Fig.~\ref{fig:2}B). However, the essential features of the solutions are the same and the generation number ($G$) dependence of the above four quantities is also very similar. Figure~\ref{fig:2}D shows a number of frames from an imaginary movie visualizing how the ability matrix converges for growing $G$ to its final state for a given set of initial abilities. Related movie files are included in the Appendix.

\begin{figure}
\centering
\includegraphics[width=.99\linewidth]{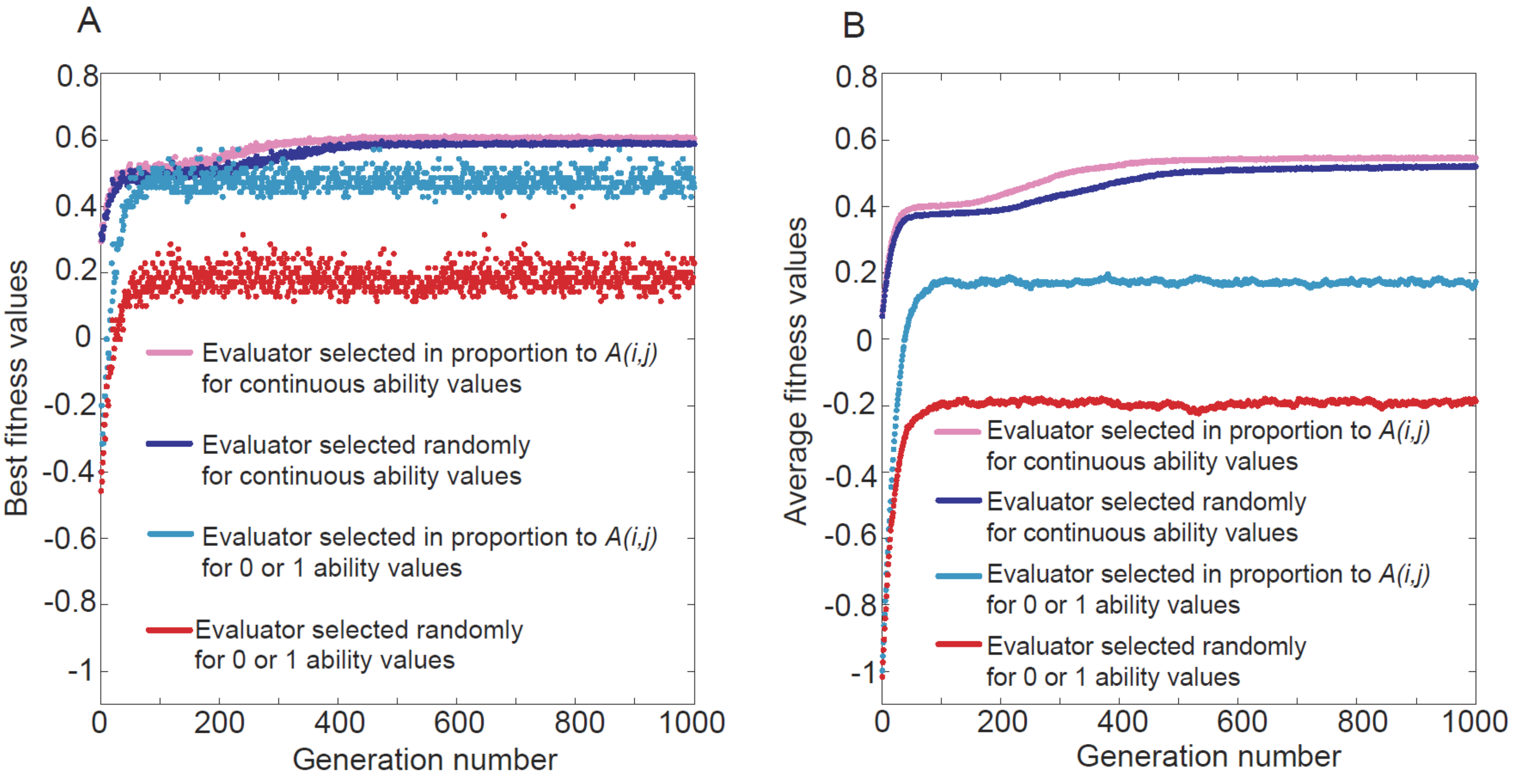}
\caption{(A) Best and (B) average performance (fitness) values as a function of $G$ obtained for the two fundamental variants for the possible $A_{ij}$: one allowing any values between 0 and one, while the second having only two possible (0 or 1) values. This is an important test to demonstrate that the intuitive, trivial choice for the abilities of the members, i.e., 1 corresponding to perfect competence while 0 corresponding to zero competence (with regard to a given sub-problem) results in less efficient groups. The reason is increased cost for $A_{ij}=1$ and the inefficient discussion phase (due to the presence of the totally incompetent members). The average fitness of the 2000 groups in a generation is significantly lower than that of the best performing ones in the binary case. The continuous distribution is much less sensitive to random perturbations than the two-valued one, the best and average performances are also very similar.}
\label{fig:3}
\end{figure}

\paragraph*{Continuous vs. two-valued}
In Fig.~\ref{fig:3} we display results obtained from an optimization of the ability matrix where the $A_{ij}$ values are, in the first case, arbitrary (continuous between 0 and 1), while in the alternative case, either $1$ (full competence) or $0$ (zero competence). In the two-valued case we expect that the trivial optimal solution is a group having $1$ specialist for each sub-problem (the same member can be a specialist for more than $1$ sub-problem, but we expect a single specialist per sub-problem). Such a solution would indeed be optimal if full knowledge was not too expensive and no discussion/evaluation took place. In reality this is not the case though since both the independent evaluation of an expert and the cost of hiring him/her are very high. This aspect of the problem can be accounted for by our cost function $f(A)$. 

Indeed, we find for realistic situations (full special knowledge is expensive and discussion can improve finding the optimal solution) that our approach results in a multiple-valued ability distribution performing better than the one constructed only from the trivial $1$ and $0$ abilities.
In short, our formalism can be used to find the appropriate strategy (choosing between hiring top specialists or implementing longer discussions). Through adding some cost for the length of the discussion phase, even the optimal discussion time can be determined.

\section*{Results based on big data analysis}
The above results are also exemplified by a number of studies on collaboration, especially on the creative groups formed by scientists working on solving increasingly complex problems. At a very recent meeting \cite{wuchty2007increasing} on interdisciplinary science it was concluded that productive interdisciplinary researchers have a deep knowledge of at least one field but also a working awareness of others. Or, in other words, during broad collaborations individuals' breadth is as important as depth of knowledge in collective decision-making. In fact, Uzzi and collaborators have shown using huge bibliographic data sets (see \cite{wuchty2007increasing,uzzi2013atypical}) that papers of high impact tend to be produced by larger collaborations involving a broader wealth of knowledge. 

It is highly non-trivial to test our theory against observations since the quantities we use are very rarely available. Still, an analysis based on a huge database (Web of Science - WoS \cite{WoS}) provides ``experimental'' evidence supporting our main theoretical result. Our method to find evidence supporting the prediction(s) of our approach was based on a very motivating remark by P. Ball \cite{ball2015communication}.

\begin{figure}
\centering
\includegraphics[width=.4\linewidth]{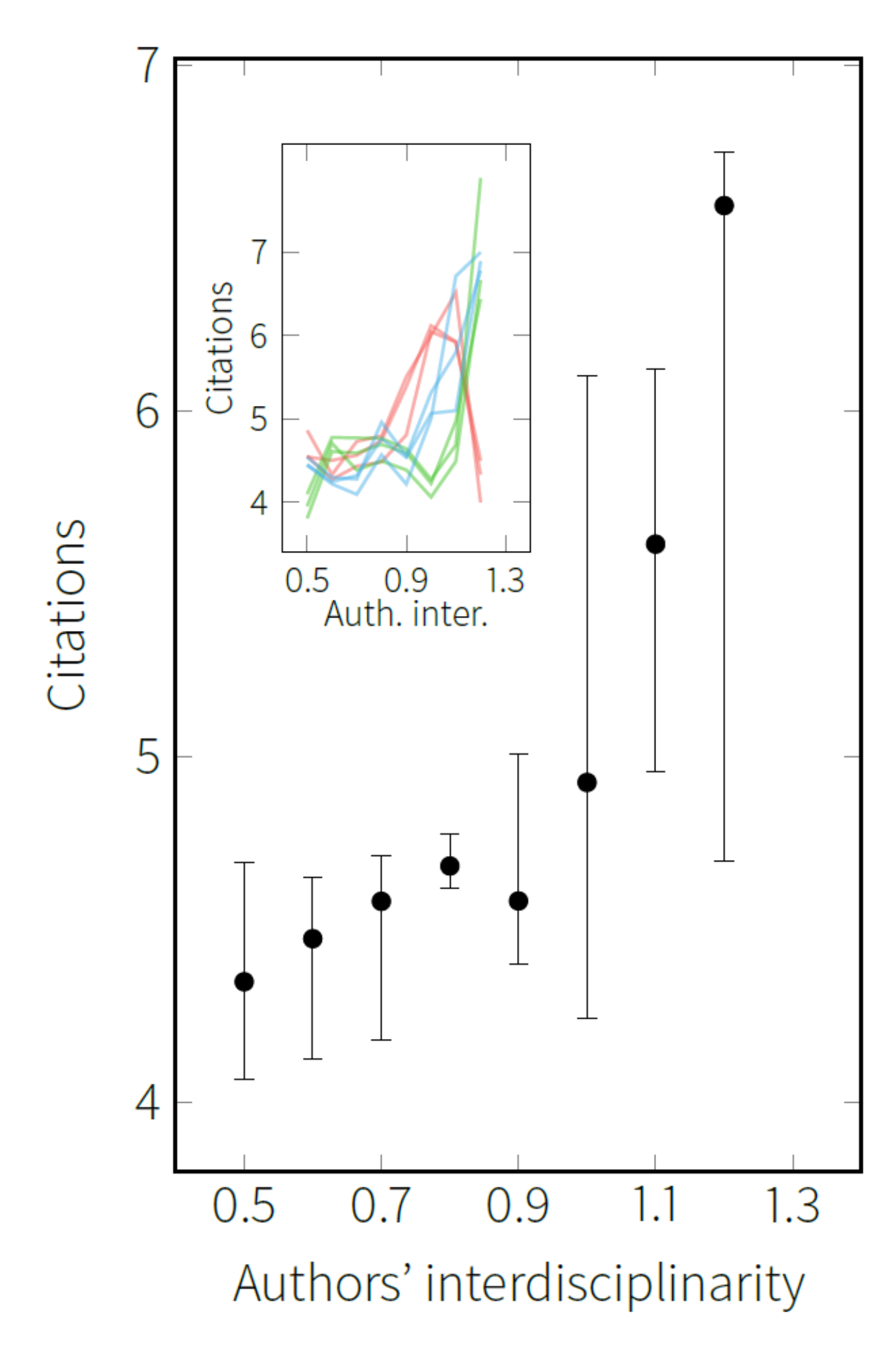}\\
\caption{Relative success of papers written by collaborating interdisciplinary scientists. Median citation number of the publications as a function of the average interdisciplinarity of the corresponding authors (measured by the average entropy of each author's publications -- averaged over the authors), error bars denote lower and upper quartiles. Data shows the median of nine trends (papers published between 1997 and 1999; for each year, annual citation count is calculated 9, 10 and 11 years post-publication). Single trends include only bins with at least 10 papers. Inset shows the different trends obtained by the three publication years and three citation years. Only papers with number of authors between 3 and 50 and authors with less than 50 publications have been considered.}
\label{fig:4}
\end{figure}

First, we selected articles published in the years 1997-1999 separately (therefore we have three sets of papers) and calculated the entropy for each article. Author's entropy was restricted to the papers published by them in the years considered (1997-1999), and only authors with at most 50 papers were considered to account for valuable contributions. Then we plot the citations of the papers as the function of the average entropy of their authors, limiting the results to papers having at least 3 and at most 50 authors. We expect highly interdisciplinary publications to show the effect of receiving high attention (and citations) only after some delay (around 10 years) to a higher extent than the less interdisciplinary ones. Therefore, for each of the four years, we considered citations from a single year with a delay of 9, 10 and 11 years. Thus, we obtained nine data sets in total, describing papers being published from 3 different years and their citations calculated with 3 different time delays. We then binned papers by their entropy in bins with 0.1 width. Results are shown in Figure~\ref{fig:4} (we ignored bins that had less than 10 papers).
As the lower and upper quartiles illustrate, for papers having high average author entropy, the citation distribution prefers higher values as well, which is also supported by the inset, where all 9 curves are shown. There is a clear trend towards more successful papers as the average author entropy increases.

\section*{Discussion}
Our formalism allows its application to more specific cases corresponding to various actual situations. It is, in some sense, the equivalent of the ``division of labor'' concept translated to the field of decision-making. It can be easily generalized to cases with various relative weights/influences assigned to the group members (depending, e.g., on their social status in an organization) when their assessment is considered. Additional future research could address further interesting questions such as, e.g., the optimal size of a group for a given number of sub-problems, the most reasonable time interval spent on discussions, the effect of ``overlapping'' problems, etc. Furthermore, the bilateral relations among the members of the group (which may be interpreted as an underlying network) can play an important role in finding the best solution.  However, the main goal of our present study, instead of demonstrating particular applications, has been to provide a general framework for further quantitative estimations of essential parameters during collective decision making concerning complex problems to be solved by multidimensional groups (as far as concerning the abilities of their members).

\begin{methods}
Three important theoretical approaches used during the research are presented in this section.

\paragraph*{Numerical procedure of optimization}
We search for the optimal ability distribution by using an improved (``simulated annealing'' type) version of the standard, so called, genetic algorithm. During this procedure, in each ``time step'' or ``generation'', two sets of $A_{ij}$ -s are randomly perturbed and merged. $A_{ij}$-s are selected for merging with a probability proportional to their level of goodness/efficiency or ``fitness'', $F$, where 

\begin{equation}\label{eq:fitn}
F = Q - C 
\end{equation}
(i.e., fitness or the efficiency, $F$, of a group is equal to quality of the solution of the complex problem minus the cost needed to obtain such an answer). Some further details and the whole optimization process algorithm are described later and in the Appendix. 

The subsection Basic assumptions represents our basic setting. Note that the concrete problem ($P$) is not specified. In addition, we have only two arbitrary parameters (the level of stochasticity, $\mathrm{Rand}$, during the second evaluation step, plus the proportion of the evaluators $X\%$). $N$ and $M$ are simple input parameters depending upon an actual situation. The description of the process may seem lengthy, however, it directly corresponds to our everyday practice during group decisions.

It is important to note that our approach is not related to the question of kin versus group selection in any way. We do not assume that the well performing groups we find evolve as a result of competition/selection between groups. Instead, we use an optimization procedure to find an efficient group which then is to be composed ``from outside'', i.e., by other people being motivated to have a group which is likely to come up with good decisions for a given class of complex problems.

We use a genetic algorithm \cite{goldberg1989genetic} to find optimal solutions because this approach is known to be effective when extreme values for a function with a relatively large number of variables is being searched. Here relatively large means numbers above $8-10$, i.e., we are looking for optima of a function which is defined in a high-dimensional space. In addition, (just like in the case of fitness landscapes or the free energy landscape for spin glasses and alike) our fitness function is likely to have a huge number of local maxima, and a single (or a class of) configuration(s) (sets of $A_{ij}$-s with maximal fitness) corresponding to a global maximum. This is why we also integrate into our approach a technique analogous to the one called ``simulated annealing'' \cite{kirkpatrick1983optimization} used for finding the minima of the free energy in the case of problems from statistical mechanics. In our case this is realized by temporarily increasing the mutation rate when the actual solution seems to converge (stops changing as a function of the generation number). Even when applying this method, one cannot be sure that in the limit of a large number of generations the absolute optimum can be reached. Thus, usually a further, quite natural, and implicitly widely used approach is taken by assuming that the pseudo-global, optimal solutions possess the same statistical features.

\paragraph*{Measuring diversity}
Since our main goal is to show the advantages of a high-dimensional, diverse ability distribution, we need a definition of diversity well suited to our approach. Our findings (visualized in Figs. \ref{fig:1} and \ref{fig:2}) indicate that the optimal $A_{ij}$ values are distributed in a way which is rather different from a Gaussian type (well defined mean and values scattered around it). Thus, for our purpose, standard deviation as a measure of diversity would not serve as a good candidate. Other widely used diversity measures (perhaps the best known of which is the Shannon information) also did not turn out to be well suited for our case. On the other hand, calculating the extent of diversity using the expression \cite{freeman1978centrality}

\begin{equation}
D = \frac{\sum \limits_{i,j} \Big[\big(\max\limits_i A_{ij}\big) - A_{ij}\Big]}{M \cdot (N-1)}
\end{equation}
gives results which differentiate among the diversity of distributions in a way being both in accord with the intuition and sensitive enough in the range determined by the actual distributions of $A_{ij}$-s throughout the simulations. 

\paragraph*{Entropy and heterogeneity}
To measure the effect of the heterogeneous ability distribution in solving a task by a group of individuals, we calculated the level of interdisciplinarity of scientific publications using the WoS database, where subject classes are assigned to each article, which in our view correspond to the different types of sub-tasks. We define the level of interdisciplinarity, $I_{\mathcal{P}}$, of a published paper by the Shannon entropy over the subject class distribution of the publications in its reference list\cite{wang2015interdisciplinarity}. More precisely, we collect all subject classes from the papers appearing among the references of the article ($\mathcal{S}_\mathrm{ref}(\mathcal{P})$) and consider the distribution obtained, thus:

\begin{equation}
I_{\mathcal{P}} = -\sum\limits_{s\in\mathcal{S}_\mathrm{ref}(\mathcal{P})} p_s \ln p_s,
\end{equation}
where $p_s$ denotes the probability of subject class $s$ in the set of subject classes based on the papers in the reference $\mathcal{S}_\mathrm{ref}(\mathcal{P})$.

Analogously, an author's interdisciplinarity is related to the average entropy of the publications this author has:

\begin{equation}
I_{a} = \langle I_{\mathcal{P}}\rangle_{\mathcal{P}\in\mathbb{P}(a)},
\end{equation}
i.e., the higher entropy corresponds to a higher level of interdisciplinarity of an author. Here $\mathbb{P}(a)$ denotes the papers of author $a$. We use the publication entropy instead of the subject class of the author's papers, since there can be authors who publish in a small number of different journals but can be still interdisciplinary. In other words, in calculating the heterogeneity of the authors' abilities, the entropy of their publications has a higher resolution and thus it provides a more accurate description of their interdisciplinarity. Finally, each paper is considered as a task, and the level of heterogeneity in the distribution of the authors' ability is defined by the average interdisciplinarity of the authors. Here we measure the success of solving the task by the number of citations the paper receives.
\end{methods}

\section*{References}
\bibliographystyle{naturemag}

\section*{Appendix}
\subsection{General presentation of the model}
On a technical level our approach can be described as optimum searching on a high-dimensional highly rugged surface. Thus it has two main components: the searching mechanism and the function defining the surface. 

For the optimum-seeking process we use a genetic algorithm enhanced with some simulated annealing-like features, which induces perturbations in the mutation rate to ensure the stability of the obtained result. An interesting feature of this approach is that it obtains results which are reachable and maintainable.

The function defining the surface includes the modeling of the given real-life problem-class and the estimation of the goodness of the actual evaluated parameters based on the constructed group-dynamical mechanism. This is called the fitness function. 

\subsection{Genetic Algorithm}
An altered version of the generic evolutionary algorithm is used to find the optimum of the fitness function. The ``individuals'' of this evolutionary process are the groups modeled through the ability matrices (a 2d array consisting of a 1d array for each member of the group, resulting in an $N\times M$ matrix, where $N$ is the number of members and $M$ the number of sub-problems). Thus the population on which the evolution acts is a collection of groups (in our case the typical population size is $K = 2000$).

The twist in our approach appears when the random point mutations are applied. Usually a predefined number of abilities from the whole population are selected, and the mutation consists in randomly increasing or decreasing their values (within a mutation amplitude range). If the mutation probability is $p_m$, the number of mutations
\begin{equation}
n_m = K\cdot N\cdot M\cdot p_m. 
\end{equation}

But in our case the pm value is not entirely fixed, it can change from generation to generation. It has a predefined normal value $p_{mn}$, which is its starting value as well. But if the difference between the averages of the fitness values in two adjacent $100$ (or $500$) generations is smaller than $0.1\%$ (or $1\%$), then the normal mutation rate ($p_{mn}$) is increased, and then annealed back to the original one (during $50$ or $100$ generations). This solution helps the algorithm to avoid being stuck in small local optima, and also ensures that the results acquired have high stability, and good resistance to small perturbations. Additionally the actual value used at each generation is defined by the following equation: 
\begin{equation}
p_m = p_{mn} \cdot (1-F), 
\end{equation}
where $F$ is the population average of the fitness value defined in the next section.

After this step is ready, only the normalization is ahead (when it is not applied, the values of the abilities appear as cost in the fitness function): here the $A$ matrices of each group from the emerging generation are normalized such that the
\begin{equation}
\langle c \cdot A_{ij}^e\rangle_{ij} = \mathrm{avg},
\end{equation} 
where $\mathrm{avg}$ is the predefined average value of the abilities.

\subsection{Fitness function}
A short description about a concrete realization of the fitness function is also included in the article, but the aim of this section is to present the most general form of it, underlining the generic mechanism of our approach, and also showing its relation to the concrete case analyzed in the simulations.

\subsection{Flowchart of the fitness function}
This function is in fact where the model of the problem-solving and solution-selection process is encoded in the whole process. The input values are the ability matrix of a given group and the return value is a real number representing the ``fitness'' of this instance. As seen in the above chart the function is totally described by giving the exact forms of the functions $F_\mathrm{proposal}$, $F_\mathrm{evaluation}$, $F_\mathrm{discussion}$, $F_\mathrm{selection}$, $F_\mathrm{aggregate}$ and $F_\mathrm{cost}$ (all of them may include stochasticity as well).

First each member of the group proposes a solution for each sub problem; these values are proportional to the members' abilities regarding the given task. In the article we considered the simplest case, where the quality of the proposed solution (for problem $j$ by member $i$, being $Q_{ij}$) is equal to the respective ability: 
\begin{equation}
F_\mathrm{proposal} \left(A_{ij}\right) = A_{ij}. 
\end{equation}

The equality ensures that the small ability values (those close to zero) do not originate from the possible noise introduced at this level.

The next step follows: each member evaluates all the solutions which were given to each sub problem, in the article we use the equation described there: 
\begin{equation}
F_\mathrm{evaluation}\left(Q_{ij}, A_{i' j}\right) = Q_{ij} \cdot  A_{i' j} + \left(1 - A_{i' j} \right) \cdot \mathrm{Rand}, 
\end{equation}
where $\mathrm{Rand}$ is a uniform random number from the interval $(0,1)$.

In step c, the discussion phase, $X\%$ of the group members ($i^{\prime\prime}$) selected with probability proportional to their ability in the respective field share their evaluations with the others ($i^\prime$), who change their own such values regarding the proposals of everybody ($i$) for each sub problem ($j$) based on this information:
\begin{equation}
F_\mathrm{discussion} \left(E_{ij}^{i'} (t), \left(E_{ij}^{i''} (t) - E_{ij}^{i'} (t)\right)\right)= E_{ij}^{i'} (t) + \frac{1}{N} \left(E_{ij}^{i''} (t) - E_{ij}^{i'} (t)\right). 
\end{equation}

So here the model supposes that everybody can be influenced in the same way by the current talker, and their opinions are changed so that the difference between their and the talkers opinion is reduced. (Note that the selection of the evaluators -- speakers -- happens proportionally to their ability values. This passage creates a situation in which the speakers are usually ``experts'' regarding the given sub-problem.)

Then the evaluations of the members are aggregated. In our case it simply means that for each sub problem the proposition which received the highest average evaluation is accepted as the solution of the group (here no hierarchy coefficient is included): 
\begin{equation}
F_\mathrm{selection}\left(E_{ij}^{i' *}, H_{ij}\right) = \max_i \sum_{i'} E_{ij}^{i' *}, 
\end{equation}
where $E_{ij}^{i' *}$ denotes the final value of $E_{ij}^{i'}$.

For calculating the final return value of the fitness function in the article the most simple and intuitive aggregation function is used: 
\begin{equation}
F_\mathrm{aggregate}\left(Q_j^{max}\right) = \langle Q_j^{max}\rangle_j.
\end{equation}

And in the simplest case (if the average ability cost does not have a predefined value – contrarily this is just a constant change in the function values) $C$ is simply a monotonous function of the ability values, but it could also include the time of decision making (which we assumed to be proportional to the number of talkers in the discussion phase) or other relevant parameters. The typical case of our approach uses 
\begin{equation}\label{eq:fcost}
F_\mathrm{cost}\left(A_{ij}, X\right) = \langle c\cdot A_{ij}^e\rangle_{ij}
\end{equation}
(with typical values $c = 4$, $e = 4$).
 
\subsection{Animation about the evolution of the ability matrix}

The animations reachable through the links present the evolution process of the ability matrix in two different bur very similar realizations of the simulation using the parameter set used in the core article as well (in the first case the ability values are represented with colors, in the second case, with colors and bars). It can be nicely seen how the specialist for each sub-problem emerges from the rest of the group, and an optimal distribution wins.

See S1 movie and S2 movie.

\subsection{Comparing the results for different ability cost coefficients and exponents}
In the case, when equation \ref{eq:fcost} holds, there are two independent parameters, namely the $c$ and $e$ constants. We present in this part the effect of modifying these values.

Firstly, we observed that the ratio of specialists in a group remained $\frac{1}{N}$ in every considered case, meaning that each sub-problem will have one specialist:
\begin{equation}
\frac{M\cdot N}{N} = M. 
\end{equation}

(This result was stable within $1\%$ of error range, where the small error could appear when the two groups -- specialists and the rest -- could not be separated perfectly.)

\begin{figure}
\centering
\includegraphics[width=.99\linewidth]{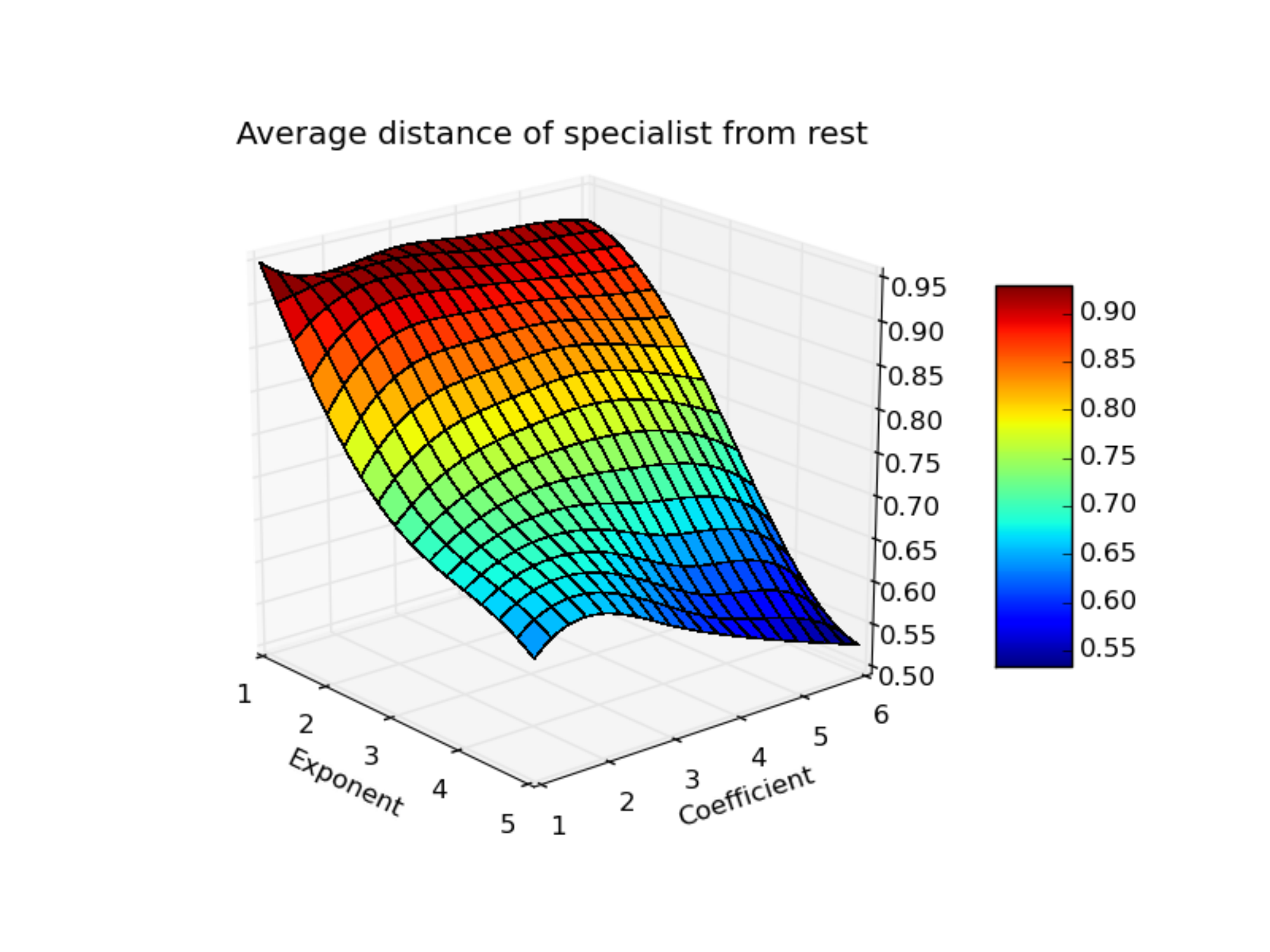}
\caption{Average distance of a specialist's ability from the rest of the members'. For the explanation of the parameters see the text.}
\label{fig:dist}
\end{figure}

Secondly, we considered the average distance between the ability of the specialists (regarding the sub-field in which they are specialists), and the knowledge of the rest of the group. (This measure is in fact a synonym of the diversity presented in the method section of the main text). The results of this inquiry are presented in Figure \ref{fig:dist}.

This surface plot makes it clear, that as the cost of outstanding knowledge increases (as the $e$ and $c$ values get higher, the difference between the cost of $0.5$ and $1.0$ ability values gets emphasized), the optimal difference between the specialist and the other members gets smaller.

Clearly this is just an example from the huge range of possible uses of the model, and unforeseeable range of its applications. 

Another outcome of this approach (of changing the two parameters of the ability cost) shows the stability of the outcome, as the results in all cases are very similar, and basically the difference between them is just the mean and standard deviation of the two peaks in the ability values histogram representing the specialists and the rest.


\section*{Acknowledgements}
The authors are grateful for the helpful comments by P. Ball and I. Scheuring.


\end{document}